\documentclass[showpacs,twocolumn,preprintnumbers,amsmath,amssymb,aps,prb,superscriptaddress]{revtex4}

\usepackage[sort&compress]{natbib}
\usepackage{graphicx}
\usepackage{dcolumn}
\usepackage{bm}

\newcommand{\cobaltite}{Co$_3$O$_4$}
\newcommand{\spinel}{MgAl$_2$O$_4$}

\begin{document}


\title{Exchange bias and interface electronic structure in Ni/{C}o$_3${O}$_4$(011)}

\author{C. A. F. Vaz}
\email[Corresponding author. Email: ]{carlos.vaz@cantab.net}%
\affiliation{Department of Applied Physics, Yale University, New Haven,
Connecticut 06520}%
\affiliation{Center for Research on Interface Structures and
Phenomena (CRISP), Yale University, New Haven, Connecticut 06520}%

\author{E.~I. Altman}
\affiliation{Department of Chemical Engineering, Yale University,
New Haven, Connecticut 06520}%
\affiliation{Center for Research on Interface Structures and
Phenomena (CRISP), Yale University, New Haven, Connecticut 06520}%

\author{V.~E. Henrich}
\affiliation{Department of Applied Physics, Yale University, New
Haven, Connecticut 06520}%
\affiliation{Center for Research on Interface Structures and
Phenomena (CRISP), Yale University, New Haven, Connecticut 06520}%

\date{\today}

\begin{abstract}
A detailed study of the exchange bias effect and the interfacial
electronic structure in Ni/\cobaltite(011) is reported. Large
exchange anisotropies are observed at low temperatures, and the
exchange bias effect persists to temperatures well above the
N{\'e}el temperature of bulk \cobaltite, of about 40 K: to $\sim$80
K for Ni films deposited on well ordered oxide surfaces, and
$\sim$150 K for Ni films deposited on rougher \cobaltite\ surfaces.
Photoelectron spectroscopy measurements as a function of Ni
thickness show that Co reduction and Ni oxidation occur over an
extended interfacial region. We conclude that the exchange bias
observed in Ni/\cobaltite, and in similar ferromagnetic
metallic/\cobaltite\ systems, is not intrinsic to \cobaltite\ but
rather due to the formation of CoO at the interface.
\end{abstract}




\maketitle

\section{Introduction}

The exchange anisotropy, first reported by Meiklejohn and
Bean\cite{MB56} for surface-oxidized Co nanoparticles, is manifested
by a horizontal shift in the hysteresis loop of magnetic systems in
contact with an antiferromagnet, an interface phenomenon explained
by exchange coupling between the spins at the interface between the
two materials.\cite{Meiklejohn62} Although studied for many decades
now, a full understanding of exchange bias at the microscopic level
remains
challenging,\cite{BT99,NS99,Stamps00,Kiwi01,Binek03,RZ07,SKH+08}
largely because of the difficulty in resolving the interface spin
and the electronic structure between the two materials. Several
microscopic models have been proposed to explain the exchange
anisotropy and the small exchange bias observed experimentally {\it
vis {\`a} vis} the strength of the spin-spin exchange coupling. The
latter effect is generally attributed to interface roughness,
antiferromagnetic domains, and steps at the interface, which lead to
a reduction in the effective exchange
coupling.\cite{BT99,NS99,Stamps00,ORS+01,Kiwi01,OSN+03,Binek03,RZ07,SKH+08}
Hence, a detailed knowledge of the interface structure is key to
understanding the mechanisms responsible for the strength of the
exchange anisotropy observed between different compounds.

Cobalt, like most 3d transition metal elements, can exist in more
than one oxidation state. Of the two stable cobalt oxides, the mixed
valence compound, Co$^{2+}$Co$_2^{3+}$O$_4$, crystallizes in the
spinel structure, with a lattice constant of $a=8.086$ \AA\ at room
temperature,\cite{PBBC80} while the CoO phase crystallizes in the
rock salt structure ($a=4.260$ \AA).\cite{Lide06} Both oxides are
antiferromagnetic at low temperatures, with N\'eel temperatures of
approximately 40 and 290 K,
respectively.\cite{Roth64,SH76,Bizette46,Bizette51,Blanchetais51,Roth58,SSW51,JRB+01}
While the critical temperature in these materials is too low for
practical devices, their well defined and localized spin structures
make them model systems for the study of exchange anisotropy. In
this context, our recent demonstration of the growth of high quality
crystalline \cobaltite(011) thin films on \spinel(011)
substrates,\cite{VWA+09} and the recent reports of enhanced
temperatures for the onset of exchange bias in
Ni$_{80}$Fe$_{20}$/\cobaltite\
heterostructures,\cite{LLT05,LLTG05,LLGS06,LLG+07} have motivated us
to examine the exchange bias in well characterized \cobaltite(011)
films, and to probe the electronic structure of the
metal/\cobaltite\ interface. We chose Ni as the ferromagnetic layer
in order to facilitate the electronic characterization of the
interface. Ni has an oxygen affinity similar to that of
Co,\cite{Reed71} and when exposed to air it forms a surface oxide
layer $\sim$3 \AA\ thick that acts as a passivation
layer.\cite{Holloway81,ZHS98,ONNK07,SP08} Here, we report the
observation of large exchange bias in the Ni/\cobaltite\ system,
which persists to temperatures well above the N{\'e}el temperature
of \cobaltite; the effect is stronger and extends to higher
temperatures in systems exhibiting greater interface roughness. From
photoelectron spectroscopy, we show that the deposition of the Ni
layer leads to a reduction of the \cobaltite\ over a few atomic
layers and to Ni oxidation at the interface between these materials.
A CoO/NiO interfacial region is found to form between the
\cobaltite\ and Ni films. Hence, we conclude that the observed
exchange bias is not intrinsic to \cobaltite\ but instead to CoO,
mediated by a NiO interfacial layer.

\section{Sample growth}

\spinel(011) single crystals were used as substrates for the growth
of \cobaltite\ because of the small lattice mismatch of --0.05\% and
good thermal and chemical stability. \spinel\ has the same crystal
structure as \cobaltite, which should preclude the formation of
antiphase boundaries that originate when lower symmetry structures
are grown on higher symmetry surfaces. Also, the direction
perpendicular to the (011) surface is characterized by having a
repeat period of four atomic planes as opposed to the eight-period
repeat along the [100] direction or the eighteen-period repeat of
the [111] direction.\cite{VHAA09} It therefore should be less
susceptible to stacking faults and antiphase boundary formation,
leading to fewer defects in the film.\cite{VWA+09}

The samples in this study were grown by molecular beam epitaxy in a
dual ultrahigh vacuum (UHV) system comprising a growth chamber (base
pressure $\sim 1\times 10^{-9}$ mbar) equipped with a reflection
high energy electron diffraction (RHEED) system, and an analysis
chamber (base pressure $\sim 3\times 10^{-10}$ mbar) for low energy
electron diffraction (LEED), x-ray photoemission (XPS) and Auger
electron spectroscopy (AES). The systems are interconnected by a
gate valve, allowing sample transfer under UHV. In this work, the
XPS spectra were obtained using the Mg K$_\alpha$ line ($h\nu =
1253.6$ eV) of a double anode x-ray source and a double pass
cylinder mirror analyzer ($\Phi$ 15-255G) set at a pass energy of 25
eV (energy resolution of about 0.8 eV). Further {\it ex situ}
characterization included x-ray diffraction and reflectometry (Cu
K$_\alpha$ line, using a Shidmazu diffractometer set in the parallel
beam geometry) and SQUID magnetometry (Quantum Design). Several
Ni/\cobaltite/\spinel(011) samples were prepared and characterized
for this study. In addition, two extra structures were grown: a
Ni/SiO$_x$/Si(111) film, to determine the presence of exchange bias
due to Ni surface oxidation; and a Ni/CoO/MgO(001) sample, to obtain
a CoO XPS spectrum and to determine the exchange anisotropy in
Ni/CoO. The growth procedures are distinct for the different types
of samples, as discussed in detail next.

\begin{figure*}[t!bh]
\begin{centering}
\includegraphics*[width=12.8cm]{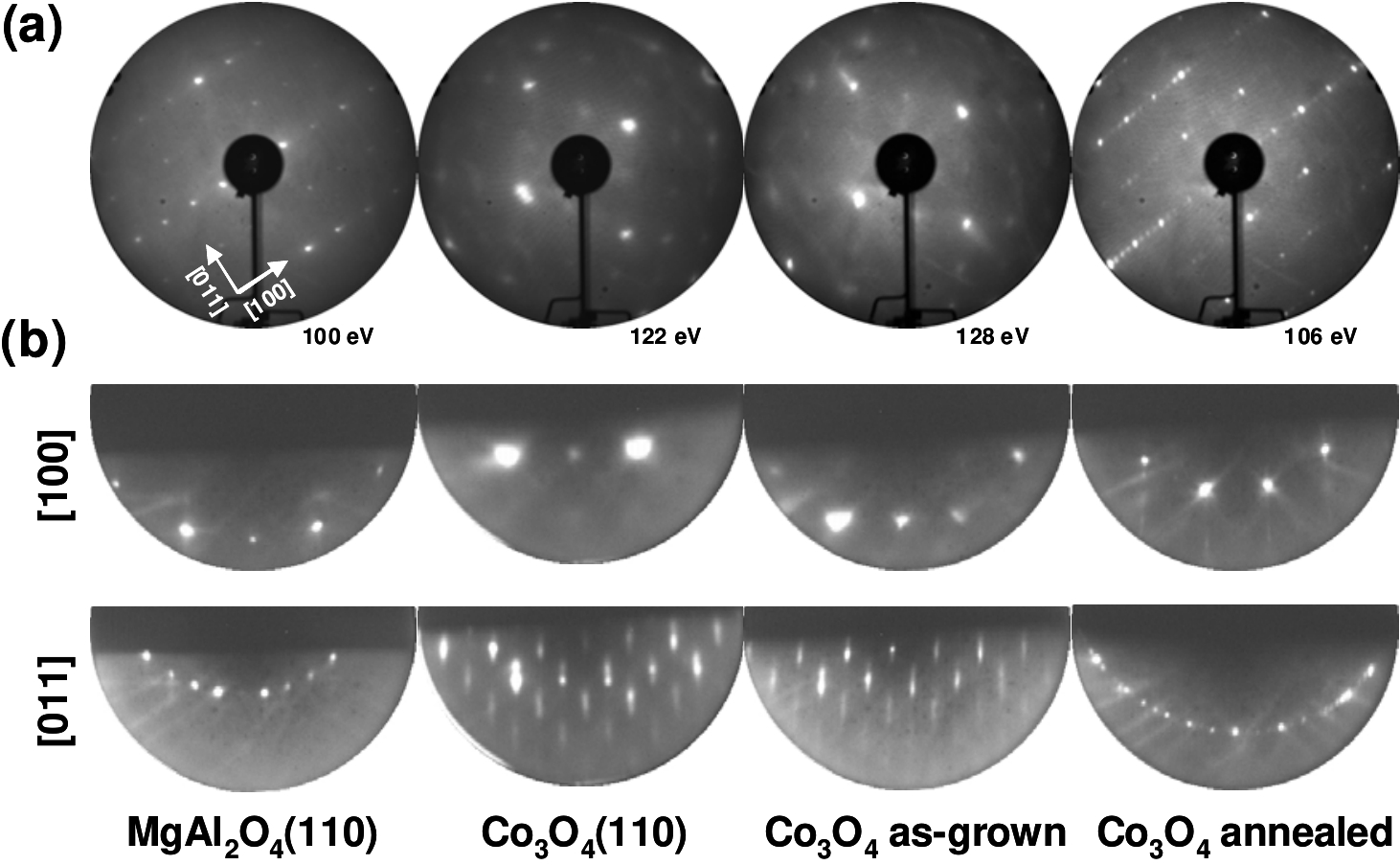}
\caption{(a) Low energy electron diffraction (LEED) patterns of the
\spinel\ and \cobaltite(011) films immediately after growth, and of
the as-grown and annealed films before Ni deposition. The crystal
orientation, inferred from the LEED pattern and confirmed by Laue
diffraction, is shown in the inset. (b) Corresponding reflection
high energy electron diffraction (RHEED) patterns, along different
azimuths (parallel to the electron beam, set at a grazing angle of
incidence). The electron beam energy was 10 keV for the
\spinel\ patterns and 11 keV for the \cobaltite\ patterns.}
\label{fig:Co3O4_ED}
\end{centering}
\end{figure*}

The \cobaltite\ samples were grown using the procedure described in
Ref.~\onlinecite{VWA+09}, with the substrate held at 570 K during
growth. The high quality of the \spinel(011) surface crystallinity
was confirmed by LEED and RHEED, which display patterns
characteristic of highly ordered surfaces, as shown in
Fig.~\ref{fig:Co3O4_ED}. The nominal \cobaltite\ thickness values,
in the 30-40 nm range, are found to be in good agreement with the
values obtained from x-ray reflectivity measurements. LEED and RHEED
patterns typical of the \cobaltite\ films after growth are shown in
Fig.~\ref{fig:Co3O4_ED}. Compared with the \spinel(011) LEED
patterns, the diffraction spots of the as-grown \cobaltite\ film are
much broader and the background is more intense, indicating that the
as-grown \cobaltite\ films have a significant amount of surface
disorder. The RHEED patterns exhibit streaky and relatively broad
diffraction spots, characteristic of a three-dimensional growth
mode. However, along the [100] direction, the diffraction spots are
found to lie in a Laue arc, indicative of a relatively smooth
surface; this is in agreement with earlier results showing a
unidirectional surface roughness morphology along this direction,
with an average surface roughness of $\sim$2-3 nm as determined from
atomic force microscopy and transmission electron
microscopy.\cite{VWA+09} After the \cobaltite\ growth and {\it in
situ} characterization, the sample was cleaved {\it ex situ}; one
half was then annealed at atmospheric pressure in flowing O$_2$ (900
scc/min) at 820 K for 12 h; this temperature and oxygen pressure
favor the formation of \cobaltite\ over CoO.\cite{TK55,KY81a,OS92a}
Both as-grown and annealed samples were then returned to the growth
system for Ni deposition. The \cobaltite\ films were first cleaned
in an O plasma at 770 K for 30 min, and in one instance cooled to
below 370 K under the O plasma. The surfaces of both as-grown and
annealed films were then characterized {\it in situ} prior to the Ni
film deposition; the corresponding diffraction patterns are labeled
in Fig. 1 as ``as-grown'' and ``annealed.''

The annealing process induces significant transformations in the
film surface structure, as shown by different and much sharper LEED
and RHEED patterns in Fig.~\ref{fig:Co3O4_ED}. In one instance, we
observe the presence of a ($4\times 1$) surface reconstruction along
the [100] direction, which we relate to the particular surface
treatment during the O-plasma cleaning procedure, i.e., cooling
below 370 K in O plasma [only ($1\times 1$) patterns were observed
previously]. One possible explanation is the presence of adsorbed
oxygen. Monitoring the LEED pattern evolution with increasing sample
temperature in UHV ($3\times 10^{-10}$ mbar) shows no changes in the
LEED pattern up to 570 K; no changes in the XPS spectra are found
after this process, indicating that the \cobaltite\ films are stable
in UHV up to that temperature. The related Fe$_3$O$_4$(110) surface
has been found to reconstruct in a ($3\times 1$)
structure,\cite{JBK95,OMT+98,MSJ+06} while a ($4\times 1$)
reconstruction has been found for the (001)
surface;\cite{AKD+97,CMJ+04} this extra periodicity in the lattice
has been attributed to the presence of Mg or Ca surface impurities
that have migrated from the bulk or from the substrate upon
annealing.\cite{AKD+97,CMJ+04,MSJ+06} Although no impurities were
detected in the XPS survey scans, a more detailed study is required
before ruling out the effect of impurities as the origin of the
observed surface reconstruction. One striking feature of the LEED
patterns is the different symmetry exhibited by the as-grown and
annealed films: while the former exhibit an oblique unit cell, the
latter are found to exhibit a rectangular unit cell. The LEED
patterns are compatible with the two possible bulk terminations of
the (011) plane of the spinel structure: type A for the annealed
films, with a Co$^{2+}_2$Co$^{3+}_2$O$_4$ nominal stoichiometry
exhibiting a rectangular unit cell, and type B for the as-grown
films, with a Co$^{3+}_2$O$_4$ nominal stoichiometry and an oblique
unit cell.\cite{VWA+09} The XPS spectra of both as-grown and
annealed films are identical and characteristic of \cobaltite.

\begin{figure}[t!bh]
\begin{centering}
\includegraphics*[width=8.5cm]{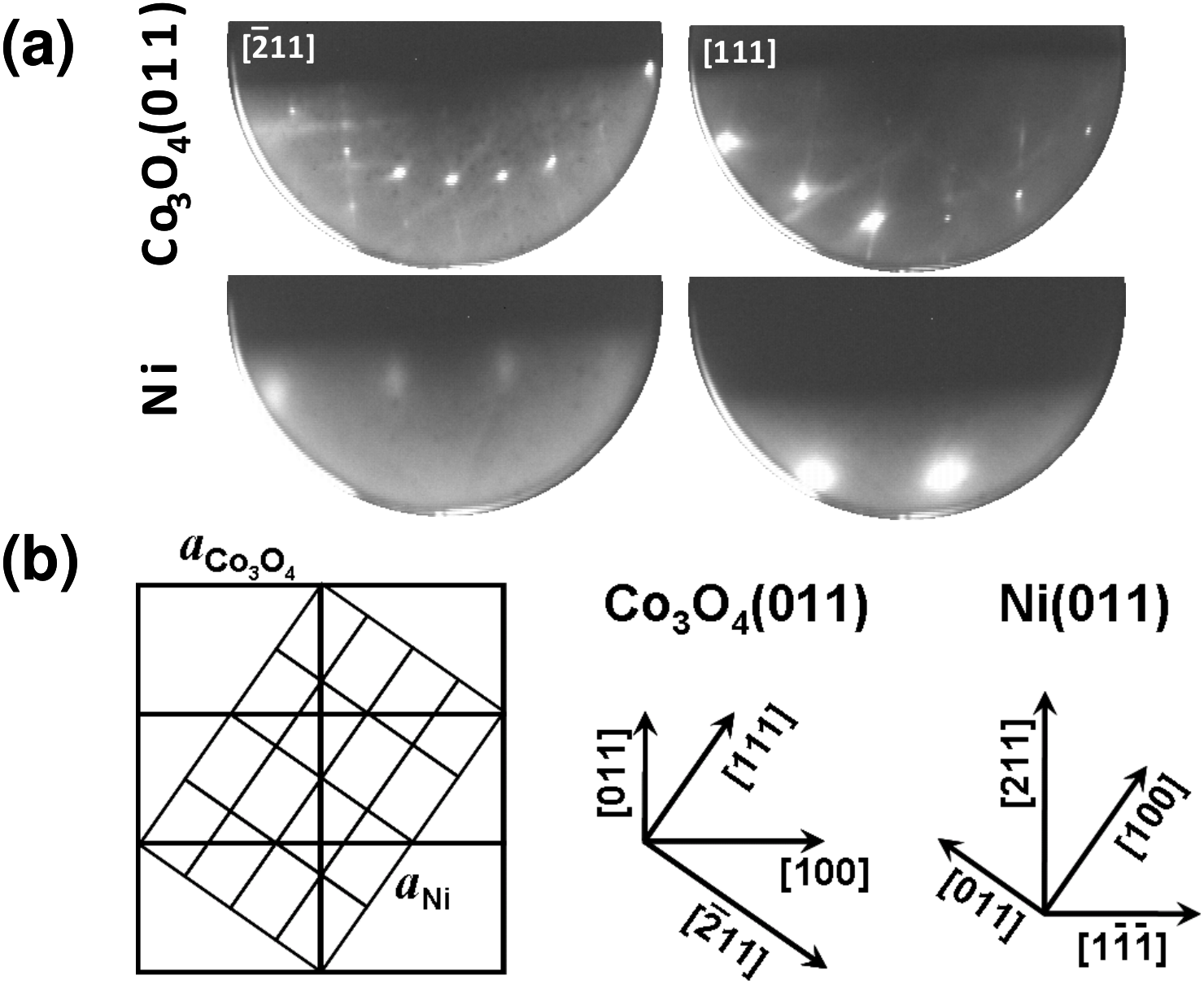}
\caption{(a) \cobaltite\ and Ni RHEED patterns along the [111] and
[$\bar{2}$11] azimuths of the annealed \cobaltite(011) film (at 11
keV incident electron beam energy). (b) Schematic of the epitaxial
relationship between [110]-oriented Ni and \cobaltite.}
\label{fig:Ni_RHEED}
\end{centering}
\end{figure}

The Ni deposition (from an effusion cell) was carried out with the
substrate held at ambient temperature; the system pressure during
evaporation was $1\times 10^{-7}$ mbar, with an evaporation rate
$\sim$1 \AA/min. We find that the Ni film grows single crystalline,
albeit with very broad RHEED patterns, as shown in
Fig.~\ref{fig:Ni_RHEED}(a), that indicate a large density of
crystalline imperfections (surface roughness and/or mosaicity), with
\cobaltite(011)[111]$\parallel$Ni(011)[100]
[Fig.~\ref{fig:Ni_RHEED}(b)] or the twin orientation
\cobaltite(011)[1$\bar{1}\bar{1}$]$\parallel$Ni(011)[100]. The above
relationship gives $\sqrt{3}a_\mathrm{Co_3O_4} = 3.97
a_\mathrm{Ni}$, where $a_\mathrm{Ni}=3.5241$ \AA\ is the lattice
parameter of fcc Ni,\cite{Stearns86} corresponding to a lattice
misfit of --0.65\%. The XPS measurements show Ni 2p peaks typical of
metallic Ni; trace amounts of O were detected in the Ni film, as
indicated by a small O 1s peak. To study the evolution in the
electronic structure at the Ni/\cobaltite\ interface, a separate
study was carried out where the XPS  spectra were recorded as
progressively more Ni (0, 1.8, 5.3, 14, 80 \AA) was added to a 30 nm
thick \cobaltite\ film. In this case, the Ni film was deposited
immediately after the \cobaltite\ growth.

Additionally, two control samples were grown, one consisting of an 8
nm thick Ni film grown on the native oxide layer of a Si(111) wafer,
and an 8 nm Ni/36 nm CoO/MgO(001) structure. For the latter sample,
the MgO(001) single crystal was outgassed and cleaned {\it in situ}
with an O-plasma at 570 K for 30 min. Cobalt evaporation was carried
out under an O$_2$ partial pressure of $1 \times 10^{-5}$ mbar. This
leads to single crystalline CoO(001), with good RHEED and LEED
patterns (not shown) and with the correct stoichiometry, as
determined by XPS (see Fig.~\ref{fig:XPS} below). X-ray diffraction
measurements confirm the presence of the [001]-oriented CoO phase,
with an out of plane lattice constant $a_\perp = 4.28$ \AA, showing
that the film is under compressive strain. The in-plane strain can
then be estimated as $\epsilon_\parallel = -\epsilon_\perp
c_{11}/(2c_{12}) = -0.42\%$, where $c_{11}=260$ GPa and $c_{12}=145$
GPa are the elastic constants of CoO,\cite{EM92a} and
$\epsilon_\perp = (a_\perp - a)/a$, where $a$ is the equilibrium
lattice constant. The misfit strain between CoO/MgO is $\eta =
-1.31$\%, showing that the CoO film is partially relaxed.

\section{Results and discussion}

\begin{figure}[t!bh]
\begin{centering}
\includegraphics*[width=8.5cm]{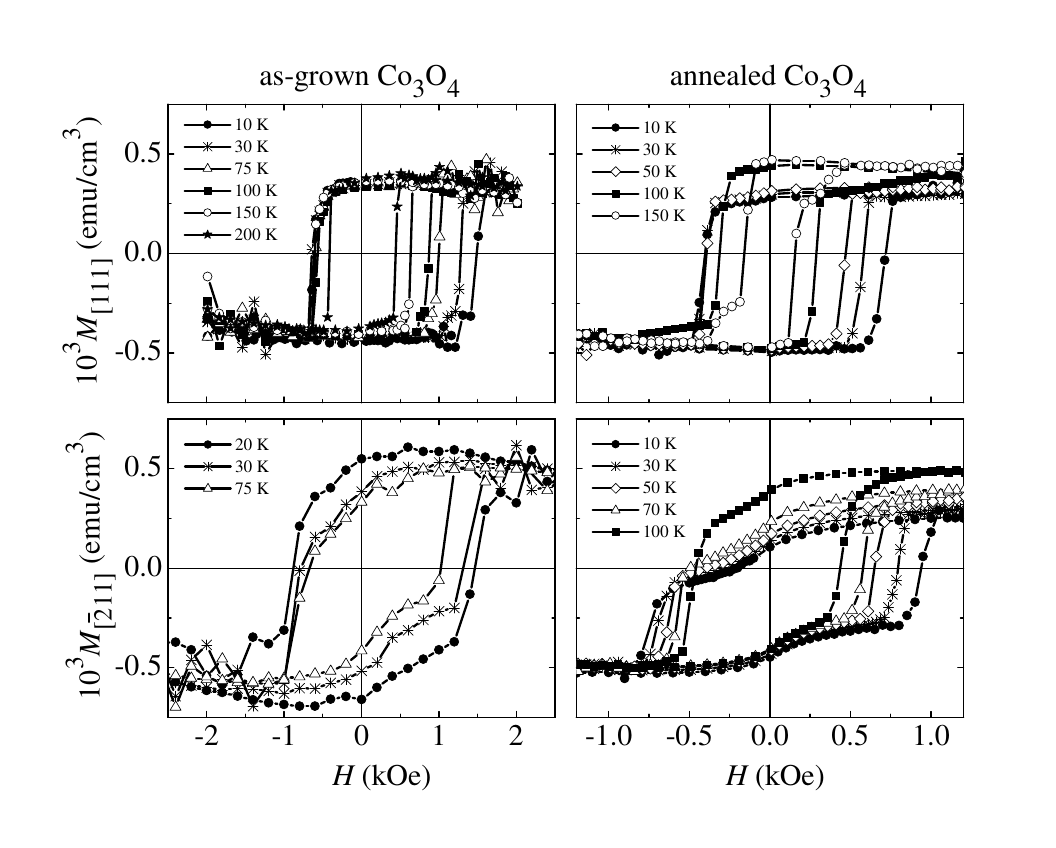}
\caption{Magnetization ($M$) versus magnetic field ($H$)
characteristics for the Ni film deposited on the as-grown (left) and
annealed (right) \cobaltite\ films for fields applied along the
[111] (top) and [$\bar{2}$11] (bottom) directions of
\cobaltite(011).} \label{fig:S36MH}
\end{centering}
\end{figure}

For the magnetic characterization, the samples were cooled from room
temperature to 10 K under a magnetic field of --50 kOe, and magnetic
hysteresis ({\it M-H}) curves were measured at increasing
temperatures in fields up to $\pm 3$ kOe. The magnetic field was
cycled once between the field range of the hysteresis measurements
at 10 K to reduce training effects (observed for the first field
cycle). For the Ni/Si(111) control sample, no shift in the {\it M-H}
curve is observed down to 10 K, ruling out the presence of exchange
bias arising from surface Ni oxidation (in agreement with the weak
exchange bias in NiO, attributed to its low magnetocrystalline
anisotropy).\cite{BT99,NS99} The magnetic measurements were
performed on several sets of samples and yielded consistent results;
in the following, we present the results for the sample set whose
LEED and RHEED patterns are shown in Fig.~\ref{fig:Co3O4_ED}. The
direction of the applied magnetic field was along the [111] and the
[$\bar{2}$11] directions of the \cobaltite(011) (the crystal
orientation was confirmed by Laue diffraction). Representative
magnetic hysteresis curves for the as-grown and annealed \cobaltite\
films along these two directions are shown in Fig.~\ref{fig:S36MH}.

\begin{figure}[t!bh]
\begin{centering}
\includegraphics*[width=8.5cm]{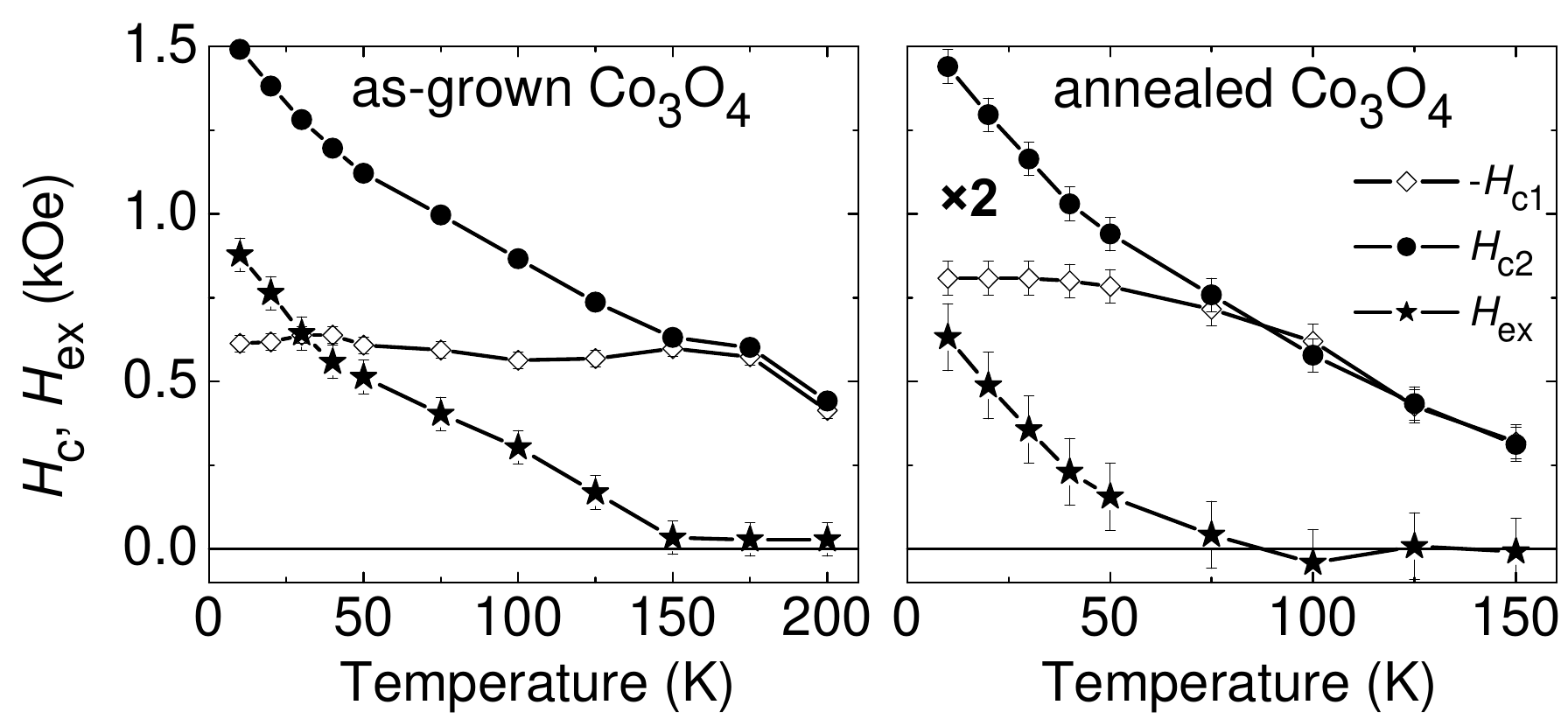}
\caption{Variation of the coercive field ($H_c$) and exchange
anisotropy ($H_{ex}$) for the as-grown and the annealed
\cobaltite(011) films along the [111] direction. The data on the
graph on the right have been multiplied by a factor of 2.}
\label{fig:S36Hex}
\end{centering}
\end{figure}

Several observations are pertinent: (i) a large shift is present in
the {\it M-H} curves in both the as-grown and annealed \cobaltite\
samples (exchange anisotropy), being significantly larger for the
as-grown samples by about a factor of two (e.g., 880 Oe for the
as-grown sample at 10~K compared to 400 Oe for the annealed sample);
(ii) the exchange anisotropy is present at temperatures well above
the N{\'e}el temperature of \cobaltite, with the as-grown films
exhibiting exchange bias up to higher temperatures than the annealed
films; (iii) a strong directional dependence of the shape of the
hysteresis curves is found, which is distinct for the as-grown and
annealed samples, and which persists at temperatures above which no
shift in the curves is observed, indicating that it originates from
magnetic anisotropies in the Ni film; and (iv) for the annealed
samples, the magnetization at positive fields does not reach
saturation when exchange anisotropy is present. The saturation
magnetization of the Ni films, $M_\mathrm{Ni} \sim$ 500 emu/cm$^3$,
is close to the bulk value of 510 emu/cm$^3$.\cite{VBL08} The
variation of the coercive field and exchange bias as a function of
temperature for the as-grown and annealed \cobaltite\ films along
the [111] direction are plotted in Fig.~\ref{fig:S36Hex}, showing
that the exchange bias sets in at around 80 K for the annealed
films, and at around 150 K for the as-grown films. This result shows
that the surface morphology of the \cobaltite\ film has a
significant influence on the exchange anisotropy, acting to enhance
it. Such an effect has been observed for single crystalline
antiferromagnetic systems.\cite{MGS95,BT99,NS99} The interface
energy, given by $\Delta E = M_\mathrm{Ni}t_\mathrm{Ni}H_{ex}$ where
$t_\mathrm{Ni}$ is the Ni film thickness and $H_{ex}$ is the
exchange bias field, is $\sim$0.35 erg/cm$^2$ for the as-grown
sample and $\sim$0.13 erg/cm$^2$ for the annealed sample at 10 K;
these values are large in comparison with those typical for
antiferromagnetic films.\cite{NS99}

The presence of a strong uniaxial magnetic anisotropy in the Ni film
agrees with the observation in RHEED and LEED of a [011]-oriented Ni
film grown on annealed \cobaltite. The magnetization lies in-plane,
with the easy magnetization axis along the \cobaltite(011)[100]
direction, which coincides with the [111] direction of Ni(011); this
is also the easy magnetization direction of bulk fcc Ni.\cite{VBL08}
However, the Ni[100] direction, which is a hard magnetization axis
in bulk Ni, is a relatively easy magnetization direction in the Ni
films, while the Ni[110] direction is a hard axis. We attribute this
behavior to a magnetoelastic contribution to the magnetic anisotropy
arising from epitaxial strain in the Ni film. This energy
contribution can be calculated within the linear theory of
elasticity, giving $E_\mathrm{me} = B\epsilon_{11}\sin^2\theta$ to
leading order, where $\epsilon_{11}$ is the in-plane strain,
$\theta$ is the angle of the in-plane magnetization vector with
respect to the Ni[100] direction, and $B$ is the effective
magnetoelastic coupling coefficient given by
\begin{equation}
B = \frac{3}{2}[(c_{11}-c_{12})\lambda_{100} + c_{44}\lambda_{111}]
\frac{c_{11}+2c_{12}}{c_{11} + c_{12} + 2c_{44}},
\end{equation}
where $c$ and $\lambda$ are the elastic and magnetostriction
coefficients, respectively. [This expression differs from that given
by Kuriki\cite{KM73,Kuriki76,Kuriki76a} in that we constrain the
magnetization to lie in the (011) plane.] Using the tabulated values
of $c_{11}=2.481$, $c_{12}=1.549$, $c_{44} = 1.242$ (units of
$10^{12}$ dyne/cm$^2$),\cite{Lide95} $\lambda_{100} = -65.3 \times
10^{-6}$, $\lambda_{111} = -27.7 \times 10^{-6}$
(Ref.~\onlinecite{Wijn91}), one obtains $B=-1.22\times 10^{8}$
erg/cm$^3$. For a fully strained Ni film ($\epsilon_{11}=-0.65$\%),
the resulting magnetoelastic anisotropy is $K_\mathrm{me} =
\mbox{+}8.0\times 10^5$ erg/cm$^3$; for comparison, the bulk
magnetocrystalline anisotropy of Ni is\cite{VBL08} $K_1 = -5.6
\times 10^{4}$ erg/cm$^3$. Since we expect the Ni film to be
partially relaxed, the magnetoelastic energy is likely to be
comparable to the intrinsic magnetocrystalline anisotropy.
Importantly, the sign of $K_\mathrm{me}$ favors $\theta=0$, i.e.,
the magnetization pointing along the Ni[100] direction,  which
agrees with the data shown in Fig.~\ref{fig:S36MH}. The magnetic
switching behavior can then be understood in terms of the local
energy minima present in the Ni(011) plane. In particular, the fact
that the magnetization does not reach saturation at positive fields
in the annealed \cobaltite\ films is explained by the switching of
the magnetization from the Ni[$\bar{1}$00] direction (parallel to
the applied magnetic field and to the exchange bias direction) to a
direction near Ni[111], which gives a magnetization projection along
the magnetic field of $\sim \cos(54.7^\mathrm{o}) = 1/\sqrt{3} =
0.58$. For the as-grown \cobaltite\ samples, an additional magnetic
anisotropy contribution arises from the presence of a directional
roughness via magnetic dipolar interaction (shape
anisotropy).\cite{Schlomann70,SJ70,Bruno88,Bruno88a,AM99,WKL+01,VSB07}
The observed coercive fields are large (and larger for the as-grown
\cobaltite\ film), suggesting the presence of strong pinning due to
epitaxial strain fields or to exchange coupling with the
antiferromagnetic spins. Magnetic measurements on the Ni/CoO(001)
control sample show the presence of exchange bias at 20 K of 150 Oe,
much smaller in magnitude than that observed in the Ni/\cobaltite\
system at that temperature; no exchange bias was present at 250 K.

\begin{figure}[t!bh]
\begin{centering}
\includegraphics*[width=8.5cm]{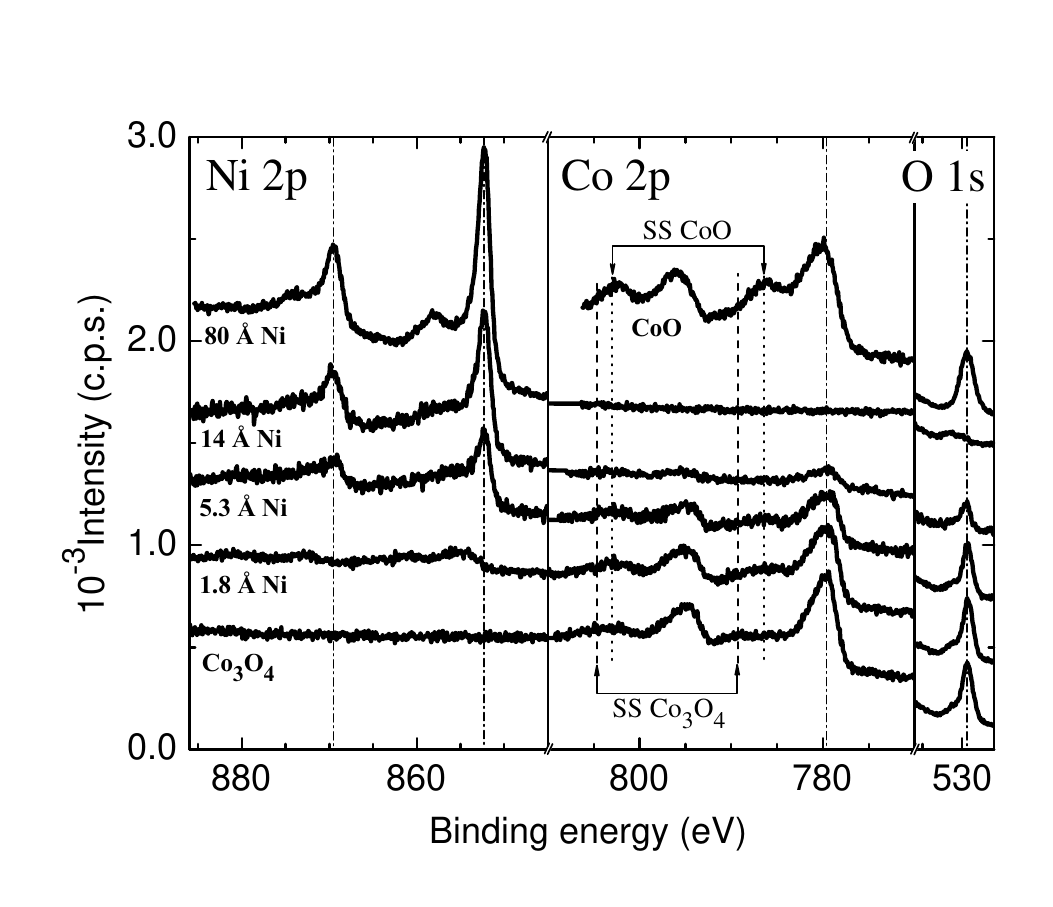}
\caption{X-ray photoelectron spectra of the Ni/\cobaltite(011)
structure for different values of the Ni film thickness at the O 1s,
Co 2p and Ni 2p edges. Also shown are the Co 2p and O 1s spectra of
the CoO(001) reference sample. Dash-dot lines indicate the positions
of the main O 1s, \cobaltite, and Ni metal peaks; dotted lines show
the positions of the satellite peaks of CoO and dashed lines the
satellite energy positions of \cobaltite. (Data have been shifted
vertically for convenient display.)} \label{fig:XPS}
\end{centering}
\end{figure}

The observation of exchange bias in Ni/\cobaltite\ is consistent
with recent reports showing the presence of exchange bias up to 150
K in ion beam-deposited Ni$_{80}$Fe$_{20}$/\cobaltite\
heterostructures\cite{LLT05,LLTG05,LLGS06,LLG+07} and up to 240 K in
Co/\cobaltite\ bilayers grown by r.f.\
sputtering.\cite{WYT+05,WTW+05,WZC+08} While in the latter studies
the authors find an extended CoO interfacial region which is
responsible for the observed temperature dependence of the exchange
bias, the former reports suggest the presence of the \cobaltite\
phase only, from transmission electron microscopy and x-ray
diffraction data. Since Ni and Co have similar oxygen affinities, a
similar interface structure is expected for Co, Ni or
Ni$_{80}$Fe$_{20}$ films in contact with \cobaltite. Therefore, to
investigate if the enhanced exchange bias effect is intrinsic to the
\cobaltite\ film or due to modifications in the interface structure,
we carried out a detailed spectroscopic characterization of the
interface by studying the evolution of the x-ray photoemission
spectra as a function of the Ni film thickness. These data are
plotted in Fig.~\ref{fig:XPS}, where we also show the CoO spectra
acquired from the Ni/CoO(001) sample. Corrections to the data
include a five-point adjacent smoothing and satellite correction;
energy shifts due to charging were corrected by aligning the O 1s
peak to the tabulated value of 529.4 eV for
\cobaltite,\cite{CBR76,CNL96,PL04} which is approximately the same
for CoO,\cite{BCR76,CNL96} NiO,\cite{WL92a} and
NiFe$_2$O$_4$.\cite{RML+97,KPBL00} The O 1s peak decreases in
amplitude with increasing Ni thickness, as expected; a small
shoulder at higher binding energies is present in the \cobaltite\
and the 1.8-14 \AA\ Ni films, but is absent in the CoO spectrum; for
the 80 \AA\ Ni film, the O 1s line lies at the energy corresponding
to the shoulder in the \cobaltite\ film, and is therefore likely to
have the same origin. The presence of such a shoulder in the O 1s
line is attributed to adsorbed
O,\cite{CBR76,JD79,KBR+84,KGBW93,GKBW95,JFEG95,CNL96,PMCL08}
although we cannot rule out the presence of adsorbed hydroxyl groups
that also give rise to a similar feature.\cite{HU77,PMCL08} Focusing
first on the Co 2p spectra, we find a significant modification of
the spectra between the 0 and 1.8 \AA\ Ni film thicknesses; the
spectrum of the 1.8 \AA\ film exhibits a significant broadening of
the main peaks and the presence of prominent satellite peaks at
binding energies lower than those characteristic of \cobaltite;
indeed, comparison with the reference CoO spectrum indicates that
these are characteristic of CoO. The conclusion is that a reduction
of the \cobaltite\ surface has occurred, leading to the presence of
a CoO-like interface layer. The same features persist at 5.3 and 14
\AA\ Ni thicknesses. The Ni 2p spectra are also revealing. The 1.8
\AA\ Ni film spectrum exhibits strong main lines that are shifted to
higher binding energies with strong satellite peaks characteristic
of Ni$^{2+}$,\cite{WL92a} which also persist up to 14 \AA\ Ni,
although a strong metallic component develops in tandem.

\begin{figure}[t!bh]
\begin{centering}
\includegraphics*[width=8.5cm]{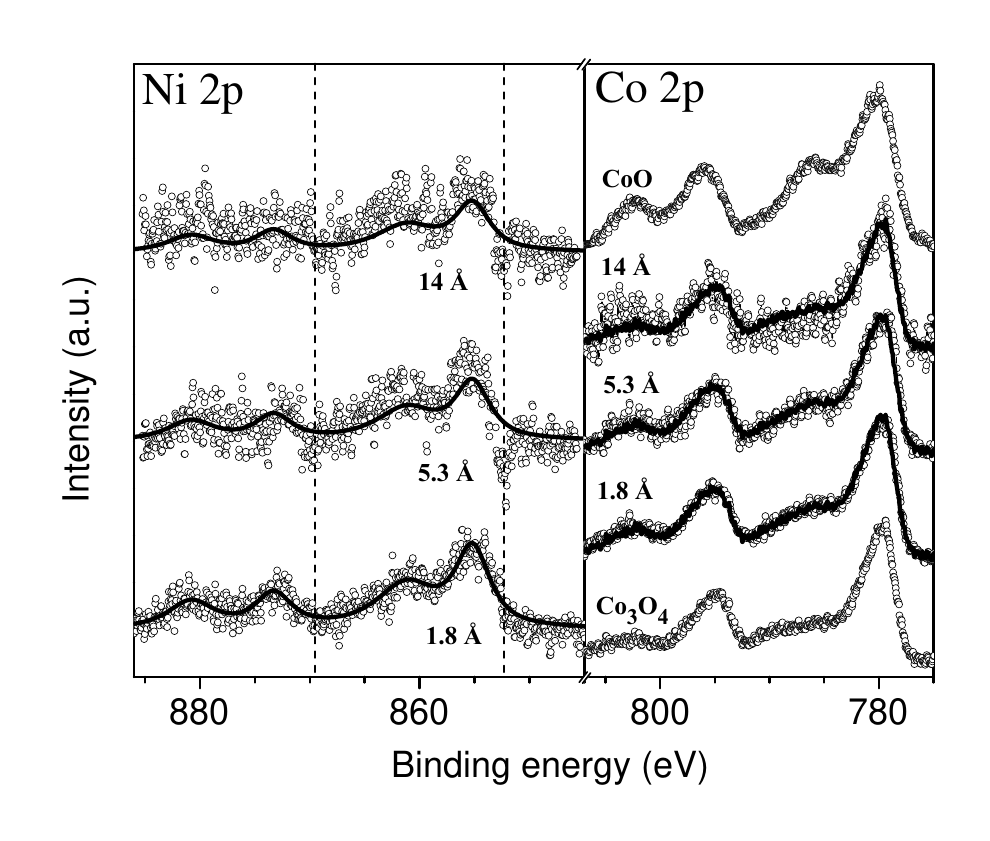}
\caption{Left panel: Ni 2p spectra of the 1.8 \AA\ Ni film (original
data) and for the 5.3 and 14 \AA\ Ni films after removing the
metallic Ni component; data are shown as circles, full lines are
guides to the eye and dashed lines indicate the position of the
metal Ni 2p main lines. Right panel: Normalized Co 2p spectra
(circles) and least squares fits to linear combinations of the CoO
and \cobaltite\ Co 2p spectra (full lines). Data have been shifted
vertically for convenient display.} \label{fig:XPS_fits}
\end{centering}
\end{figure}

Figure~\ref{fig:XPS_fits} shows a more detailed view of the Ni 2p
and Co 2p spectra for the different Ni film thicknesses, where a
linear background was subtracted from the data. The 1.8 \AA\ Ni
spectra is the same as in Fig.~\ref{fig:XPS}, plotted on an expanded
scale. One striking observation is the presence of strong satellite
peaks at 6.0 and 7.6 eV above the main peaks; the corresponding
values for NiO lie at 7.3 and 8.2 eV, respectively.\cite{FL85} Those
differences are well beyond the experimental uncertainty; the
observed satellite peak positions agree well with those for
NiCo$_2$O$_4$, 6.4 and 7.5 eV, respectively,\cite{KPBL00} which may
be attributed to the altered electronic environment of the Ni
cations or to the presence of a NiCo$_2$O$_4$ interfacial layer. For
the 5.3 and 14 \AA\ Ni 2p spectra, we use the fact that the metallic
Ni 2p$_{3/2}$ peak is well separated from the oxide Ni peak to
subtract the metal Ni contribution from the 5.3 and 14 \AA\ Ni XPS
spectra, using the scaled 80 \AA\ Ni spectrum as the reference. The
remaining signal is attributed to the Ni$^{2+}$ contribution;
despite the low signal to noise ratio, the spectra are found to have
features similar to the 1.8 \AA\ Ni spectrum. To guide the eye, we
have superimposed on the data the result of a multipeak fit to the
1.8 \AA\ Ni spectrum. The noise in the data precludes a quantitative
estimate of the Ni oxide layer, but the attenuation in the Ni oxide
signal with increasing thickness indicates that the Ni oxidation is
confined to the interface, to within 2-4 \AA, or 1-2 atomic layers.

The Co 2p spectra shown in Fig.~\ref{fig:XPS_fits} are normalized to
the 2p$_{3/2}$ peak, which removes the attenuation from the Ni
overlayer. In both this case and in that of the 2p Ni metal peak,
the scaling factors follow an exponential decay, with an attenuation
length ($\lambda$) of about 11 \AA, which agrees with the tabulated
effective attenuation lengths for the Co 2p$_{3/2}$ and Ni
2p$_{3/2}$ lines in Ni for our measurement geometry.\cite{PJ09} For
the Co 2p lines, we attempt to estimate the thickness of the CoO
layer by assuming that the Co 2p spectra of the 1.8, 5.3 and 14 \AA\
Ni thicknesses can be described by a superposition of the
\cobaltite\ and CoO bulk spectra, $I_0 = \alpha I_\mathrm{CoO} +
\beta I_\mathrm{Co_3O_4}$. The resulting fits using least squares
are shown as continuous lines in Fig.~\ref{fig:XPS_fits}. We find
$\alpha = 0.28(1)$, $\beta = 0.75(1)$ for the 1.8 \AA\ Ni film;
$\alpha = 0.30(2)$, $\beta = 0.69(2)$ for the 5.3 \AA\ Ni film; and
$\alpha = 0.11(3)$, $\beta = 0.81(3)$ for the 14 \AA\ film, where
the error bars correspond to three standard deviations. To first
approximation, the scaled Co 2p XPS signal, $I_0 =
I\exp\{+t_\mathrm{Ni}/\lambda_\mathrm{Ni}\}$ (where $I$ is the
measured intensity), is the sum of the signals originating from the
CoO and the thick \cobaltite\ layers, $ I_0 = I_\infty (1-
\exp\{-t_\mathrm{CoO}/\lambda\}) + I_\infty
\exp\{-t_\mathrm{CoO}/\lambda\}$, assuming similar effective
attenuation lengths and bulk intensities for the two Co oxides. The
intensity should be therefore approximately independent of the Ni
film thickness, with $\alpha + \beta \approx 1$, in reasonable
agreement with the fitting results given the approximations made.
For the 14 \AA\ Ni film the agreement is less good, although the fit
is also less satisfactory due to the larger scatter in the data.
(Since the signal in this case weighs the interface region more
strongly, one other possibility may be the presence of a
NiCo$_2$O$_4$ interfacial layer; the latter is characterized by a Co
2p edge spectrum with strongly suppressed satellite peaks from the
Co$^{3+}$ cations occupying an octahedral
environment,\cite{KPBL00,VPAH09} which would resemble the
\cobaltite\ spectrum more closely.) Using $\alpha = 0.3$, we
estimate $t_\mathrm{CoO} \approx 4$ \AA, using $\lambda = 11$ \AA.
The extent of the \cobaltite\ reduction may also be deduced from the
oxygen intake from the Ni interfacial layer according to Ni +
\cobaltite\ $\longrightarrow$ NiO + 3CoO. Using a Ni atomic surface
density similar to that of the Co$^{3+}$ density in \cobaltite(110),
and a Ni oxidation of 1-2 atomic layers, this results in the
formation of 3-6 CoO molecules into the cobalt oxide film,
corresponding approximately to a thickness of 4.5-9 \AA, using the
(110) interplanar distance of 1.48 \AA\ for CoO, in reasonable
agreement with the estimate obtained from the XPS data.

Based on the XPS results, we can therefore rule out a pristine
Ni/\cobaltite\ interface, and one must conclude that the exchange
bias effect observed in this system is not intrinsic to \cobaltite.
In fact, interfacial modifications in the oxidation state of CoO and
NiO in contact with 3d ferromagnetic metals are also known to
occur,\cite{ROS+01,ORS+01,YCZ+01,TMHK06,APR+07} while changes in the
oxygen stoichiometry of CoO has also been found to have a marked
influence on the exchange bias of Fe/CoO couples.\cite{NRR+07} Two
possible effects could contribute to the exchange anisotropy
observed in Ni/\cobaltite. One effect refers to the presence of the
CoO layer, which has a bulk N{\'e}el temperature of 290
K.\cite{Bizette46,Bizette51,Blanchetais51,Roth58,SSW51,JRB+01} Since
the CoO layer remains relatively thin ($\sim$4 \AA), the N{\'e}el
temperature is depressed,\cite{AC96,ATV+98,KMB+02} but would
otherwise give rise to large exchange bias at low temperatures; the
largest exchange bias effects reported in the literature have been
observed for CoO.\cite{NS99} The other effect results from the
presence of the nickel interfacial oxide layer. Ni$_x$Co$_{1-x}$O is
antiferromagnetic with a critical temperature that varies linearly
with $x$ between that of CoO and NiO, well above that of
cobaltite.\cite{Bracconi83} Since the exchange bias effect is an
interfacial effect, this layer must necessarily mediate the exchange
coupling between the CoO and the Ni film. Despite its small
thickness, it is likely to be magnetically polarized from contact
with the Ni and the CoO layers. Both effects are consistent with the
observation of exchange bias in Ni/\cobaltite\ at temperatures well
above the N{\'e}el temperature of \cobaltite.

\section{conclusions}

In conclusion, we have carried out a detailed study of the exchange
bias and of the interfacial structure in Ni/\cobaltite(011). The
exchange anisotropy is large, persists to temperatures well above
the N{\'e}el temperature of bulk \cobaltite, and is larger for the
rougher \cobaltite\ films. From photoelectron spectroscopy carried
out as a function of the Ni film thickness, the oxidation state of
the \cobaltite\ and Ni films are found to be strongly modified at
the interface, leading to the formation of a $\sim$4 \AA\ CoO layer
and of a monolayer thick interfacial NiO layer. Such modifications
in the interfacial oxidation state and electronic structure are
difficult to assess with structural characterization techniques
alone. We attribute the unusual exchange bias behavior found in
Ni/\cobaltite\ and similar metal/\cobaltite\ structures to the
presence of such a modified interface. These results emphasize how
interfacial reactions play a strong role in governing exchange bias
and thus how detailed knowledge of the interfacial structure and
electronic properties are essential to understanding these
phenomena.

\begin{acknowledgments}
The authors acknowledge financial support by the NSF through MRSEC
DMR 0520495 (CRISP).
\end{acknowledgments}


\end{document}